# The Ideal Electromechanical Oscillator System.

Osvaldo F. Schilling

Departamento de Física, Universidade Federal de Santa Catarina, Campus,

Trindade, 88040-900, Florianópolis, SC. Brazil.

email: osvaldof@mbox1.ufsc.br fax: +55-483319946

Abstract:

Oscillators and rotators are among the most important physical systems. For centuries the only known rotating systems that actually reached the limits of the ideal situation of undamped periodical motion were the planets in their orbits. Physics had to develop quantum mechanics to discover new systems that actually behaved like ideal, undamped, oscillators or rotators. However, all examples of this latter systems occur in atomic or molecular scale. The objective of the present letter is to show how the limit of ideal oscillating motion can be challenged by a man-made system. We demonstrate how a simple model electromechanical system consisting of a superconducting coil and a magnet can be made to display both mechanical and electrical undamped oscillations for certain experimental conditions. The effect might readily be attainable with the existing materials technologies and we discuss the conditions to circumvent energy losses. The result is a lossless system that might generate hundreds of Ampere of rectified electrical current by means of the periodical conversion between gravitational potential, kinetic, and magnetic energies.

*" We will start by doing the possible. As time goes on we will end up achieving even the impossible...". President Luis Inacio Lula da Silva of Brazil, during his 2002 presidential campaign.*

One of the many unattainable dreams that have challenged the human mind and ingenuity since ancient times is that of the practical implementation of the so-called perpetual motion. The actual achievement of a perpetual motion presupposes either the existence of an infinite supply of energy to promote motion in a dissipative system or the complete conservation of the initial energy of the system. From the days of Kepler and Newton until the advent of Quantum Mechanics planetary gravitational motion was the only known example of conservative rotational motion, that is, a system that would perform periodical rotational motion in a way akin to the ideal perpetual motion. With the development of atomic and molecular physics many oscillating and rotating systems in their fundamental states were found to behave like ideal quantum oscillators or rotators. The existence of discrete energy levels in the atomic scale would maintain stability of motion and conservation of energy. These however are atomic-scale systems. All other man-made macroscopic oscillating systems have however remained affected by limitations imposed by energy dissipation, so that the ideal perpetual motion has never been achieved for such systems. The development of superconductor materials in some way renewed hopes that at last a man-made system displaying complete energy conservation might eventually be designed. As Fritz London put it in the 1930s, a superconductor contains electrons in a macroscopic quantum state. The stability of a detached energy level separated by a gap from a continuum of states, and the so-called "rigidity" of the wave-function across the material give rise to perfect conductivity. A superconducting ring can be made to transport an effectively resistanceless current for an indefinite period of time. Such motion might be considered perpetual and the associated magnetic and electron kintetic energies are conserved in the motion. However such example still involves the motion of atomic-scale particles only. That is, the energy-conserving motion of some macroscopic object that would actually fulfill the ancient ideal of the perpetual motion has never been achieved by any man-made system.

As we now show, superconductivity has eventually been the key for obtaining an acceptable solution for this puzzle. In fact, although not explicit in the idea of resistanceless electronic motion, as we show below such concept bears within itself the potential for the development of a method for its own conversion into dissipationless mechanical motion. The idea is demonstrated for a simple system consisting of a rectangular superconducting coil and a magnet. To achieve such electromechanical motion conversion the superconducting coil is submitted simultaneously to a constant magnetic field and to an external force. For such conditions the coil will oscillate mechanically around an equilibrium position, and will have induced in it a supercurrent containing an alternating component of same frequency. We will discuss the conditions necessary for such system to conserve its initial energy.

Let′s consider the model experimental setup described in Figure 1. A rectangular superconducting coil of mass *m* is submitted to a magnetic field *B* from a system of magnets. It is important that the entire coil be submitted to such field, as discussed later. The ( say) lower side of size *a* must be covered by a coating of *an electrically insulating high-magnetic permeability substance*, to produce on this side of the coil a magnetic field which will be different from the field in the ( say) upper side. The coil is pulled away from the magnets by an external force *F*, which is assumed independent of time. According to Faraday's Induction Law the motion of the coil in the presence of *B* will induce in it a transport supercurrent *i*. Such current will generate a magnetic force with strength $F_{m1}= iaB_1$ in the upper side ( of length *a*) of the coil in opposition to *F,* and another force of strength $F_{m2}= iaB_2$ in the lower side. Here $B_1$ and $B_2$ are the effective magnetic fields at the position of the sides of length *a*, and the forces must not cancel each other otherwise no effect is obtained. $B_1$ should be the constant and uniform field produced by the magnet( corrected for demagnetization factors). We will assume $B_1>B_2>B_{c1}$, where $B_{c1}$ is the superconductor lower critical field. The coil will move with speed $v$ described by Newton′s Law:

$$m\, d\, v/dt = F - ia(B_1 - B_2) \tag{1}$$

We introduce the parameter $B_0 \equiv B_1 – B_2$ to simplify the notation. The displacement of the coil gives rise to an induced electromotive force $\varepsilon$, given by Faraday's Induction Law[1]:

$$\varepsilon = -\, d\Phi_m/dt - L\, di/d\, t \tag{2}$$

Here $\Phi_m$ is the magnetic flux from the external source that penetrates the rectangular area bound by the coil, and $L$ is the self-inductance of the coil. In (2), $d\Phi_m /dt = -B_0 av$. A consequence of perfect conductivity is that the electromotive force may be considered zero, so that equation (2) implies magnetic flux conservation. It is important to stress the absence of dissipative terms in (2) , a crucial consequence of perfect conductivity that will extensively be discussed later. Taking the time derivative of (1) and eliminating $di/dt$ from (2) one obtains:

$$m\, d^2 v/dt^2 = -\, (\, B_0^{\,2} a^2/L\, )\, v \tag{3}$$

This means that the coil should perform an *ideal* oscillating motion under the action of the external and magnetic forces. Assuming zero initial speed and an initial acceleration equal to $F/m$ equation (3) can be solved:

$$v(t) = F/(m\Omega)\ \sin(\Omega t) \tag{4}$$

The natural frequency of the oscillations is $\Omega = B_0a/(mL)^{1/2}$. The amplitude of the oscillating motion will be $x_0 = F/(m\Omega^2)$. It is possible then to combine (1)–(3) to obtain an equation for the current $i(t)$:

$$(B_0a/\Omega^2)\ d^2i/dt^2 = F - B_0ai \tag{5}$$

whose solution is

$$i(t) = (F/(B_0a))\ (1 - \cos(\Omega t)) \tag{6}$$

for $i(0) = di/dt(0) = 0$. From equation (6) we conclude that the supercurrent induced in the coil is already rectified. It never changes sign, and it looks like the result of submitting an alternating current of frequency $\Omega/2$ to a diode bridge rectifier. Two details make this machine different from usual superconductor motors or generators. Firstly, the motion is linear and does not involve high speeds. Secondly, levitation does not involve highly hysteretic repulsion forces against dipolar magnetic fields, but comes straight from Lorentz forces acting upon the currents in the coil. Typical figures for the speed amplitude ($v_o$), oscillation amplitude, and rectified current amplitude ($i_o = F/(B_0a)$) may be obtained. Let's take $L = 10^{-7}$ H, $B_0 = 0.3$ T, $m = 1$ kg, $F =$ 10 N, $a$= 0.1 m. For these parameters, $\Omega$= 95 rad/s, $i_o$= 333 A, $v_o$= 11 cm/s, and $x_o$= 1.1 mm. That is, a large rectified current may be obtained with low speed and frequency of oscillation, and very small displacements of the coil.

The set of eq. (1)-(6) was deduced under the assumption of the absence of energy losses. Let's discuss the conditions for such assumption to apply. First of all, it is important to stress that the predicted mechanical motion can be made perfectly frictionless with this design, since in the vertical position the motion is independent of any physical contact between the levitating coil and the magnet. If friction losses are negligible, the next issue that immediately arises is that of the possible losses related to the alternating current in the wire and to inductive coupling between the coil and resisive conductors in the surroundings. Inductive coupling between the coil and a conducting magnet like Nd-Fe-B will produce resistive eddy currents in the magnet. This source of dissipation might be eliminated by using an insulating magnet like ferrite, at the possible cost of having to work at lower fields. For the same reason the lower coil side coating must be electrically insulating. Assuming that this source of energy dissipation may be circumvented let's discuss how to avoid losses associated with the transport of ac currents by the superconductor wire[2-3]. Such losses may be classified in resistive losses and hysteretic losses. Resistive losses might arise due to partial current transportation by normal electrons. The influence of normal electrons may be avoided by working at low frequencies[3]. We note that the number of free parameters of the model allows the frequency $\Omega$ to be set, e.g., in the 10 ~ 1000 rad/s range, so that the influence of eddy current losses due to normal carriers expected already in the upper MHz range[3] may be entirely neglected. Alternating currents give rise to a resistive response from the superconducting electrons also[4]. However, well below $T_c$ such resistivity is smaller than the normal state resistivity by a factor $\approx h\Omega/\Delta$, where $\Delta$ is the energy gap[4]. Such factor is of the order of $10^{-9}$ for low frequencies, so that the resistive behavior of superconducting electrons may safely be neglected. Hysteresis losses occur whenever the flux line (FL) lattice inside a type-II superconductor is cyclically rebuilt by an oscillating magnetic field[2]. In the present case such self-fields ( ripple fields) are created by the alternating part of the current. The work of Campbell, Lowell, and others[5-7] has shown that provided the displacements $d_o$ of the FL from their pinning sites are small enough, such displacements are elastic and reversible, and *no* hysteresis losses occur. The threshold value of $d_0$ was found to be a fraction of the inter-FL spacing, as small as the coherence length ( 2 ~ 6 nm)[5]. Let $b$ represent the ripple-field amplitude at the surface of the wire, and $B$ the static magnetic field. Campbell[5] has shown that there will be no losses if $b < (\mu_0 B J_c\, d_0)^{1/2} =$ $b_0$, where $J_c$ is the critical current density and $d_0$ is given above( we must point out that demagnetization effects must be taken into account in the value for $B$ in this formula). This explains why the entire coil should be submitted to the static magnetic fields. To avoid self-field hysteresis losses the value of $B$ must be large compared to $b$, otherwise the inequality will not be satisfied. The current $i$ will flow within a surface sheath of thickness $(Bd_0/(\mu_0 J_c))^{1/2}$. There will be no losses associated with this current provided it generates a surface field $\mu_0 i/(2\pi r)$ smaller than $b_0$ . Just to give a numerical example, if a cold-drawn Nb-Ti wire of radius $r$ = 0.085 mm, and $J_c$= 5x10$^9$ A/m$^2$[8] is used to make the coil, there will be no hysteresis losses for a transport current as high as 1.5 A in such thin wire ( adopting $B$ = 1 T). A thicker cable might carry hundreds of Ampere. One might mention also flux-creep as another possible source of dissipation, but the use of a temperature of operation well below $T_c$ , and a strong-pinning material should discard the possibility of creep.

In conclusion, according to the foregoing discussion present day technology of magnet manufacturing together with superconductor materials technology may lead to a complete elimination of energy dissipation for the oscillating system analysed in this work. Therefore, a man-made electromechanical system that conserves energy ( and thus performing a perpetual motion ! ) is technically feasible.

The author wishes to thank Prof. Said Salem Sugui Jr. for his support.

Other applications may be found in: arXiv.org/abs/cond-mat/0309043.

References.

Figure caption.

Figure 1: A rectangular coil is submitted simultaneously to an external force $F$ and a magnetic field $B_1$ perpendicular to it. The lower side of the coil feels a different field $B_2$ since it is covered by a high-permeability coating. As shown in the text the predicted motion is oscillatory, with an alternating current of same frequency being induced in the coil.

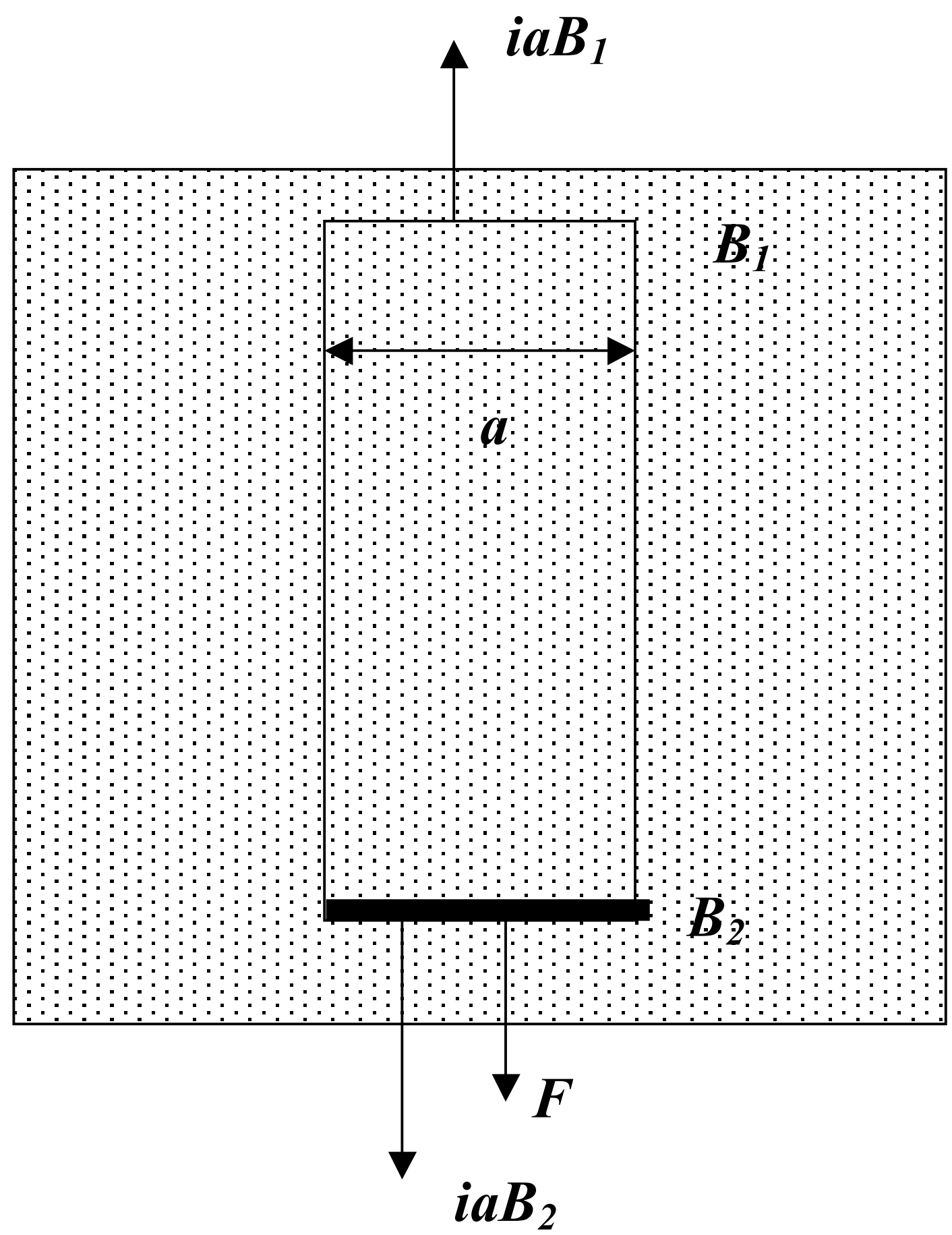

Figure 1
Schilling